# Crystal Systems' Classification of Phosphate-Based Cathode Materials Using Machine Learning for Lithium-Ion Battery

Yogesh Yadav[1], Sandeep K Yadav[2], Vivek Vijay[3], Ambesh Dixit[1,4,*]
[1]Advanced Materials and Device (A-MAD) Laboratory, Department of Physics, Indian Institute of Technology Jodhpur, Rajasthan, India, 3420230
[2]Department of Electrical Engineering, Indian Institute of Technology Jodhpur, Rajasthan, India, 342030
[3]Department of Mathematics, Indian Institute of Technology Jodhpur, Rajasthan, India, 342030
[4]Rishab Centre on Research and Innovation in Clean Energy, Indian Institute of Technology Jodhpur, Rajasthan, India, 342030
*ambesh@iitj.ac.in

## Abstract:

The physical and chemical characteristics of cathodes used in batteries are derived from the lithium-ion phosphate cathodes' crystalline arrangement, which is pivotal to the overall battery performance. Therefore, the correct prediction of the crystal system is essential to estimate the properties of cathodes. This study applies machine learning classification algorithms for predicting the crystal systems, namely monoclinic, orthorhombic, and triclinic, related to Li–P– (Mn, Fe, Co, Ni, V)–O based Phosphate cathodes. The data used in this work is extracted from the Materials Project. Feature evaluation showed that cathode properties depend on the crystal structure, and optimized classification strategies lead to better predictability. Ensemble machine learning algorithms such as Random Forest, Extremely Randomized Trees, and Gradient Boosting Machines have demonstrated the best predictive capabilities for crystal systems in the Monte Carlo cross-validation test. Additionally, sequential forward selection (SFS) is performed to identify the most critical features influencing the prediction accuracy for different machine learning models, with Volume, Band gap, and Sites as input features ensemble machine learning algo-rithms such as Random Forest (80.69%), Extremely Randomized Tree (78.96%), and Gradient Boosting Machine (80.40%) approaches lead to the maximum accuracy towards crystallographic classification with stability and the predicted materials can be the potential cathode materials for lithium ion batteries.

## I. INTRODUCTION

Lithium-ion batteries (LiBs) play a pivotal role in modern energy storage technologies, powering various applications, from portable electronics to electric vehicles [1] and grid-scale energy storage [2]. The performance, efficiency, and longevity of LiBs are significantly influenced by the properties of their cathode materials, which are often governed by their crystal structure [3]. Understanding and predicting the crystal system of cathode materials is crucial because the arrangement of atoms in a crystal lattice directly affects key characteristics of the material, such as ion diffusion pathways [4], electrical conductivity [5], and stability under operating conditions [6]. The selection and optimization of cathode materials have traditionally relied on experimental techniques, such as X-ray diffraction (XRD) [7], and computational methods, like density functional theory (DFT) [8], to determine their crystal systems. Although these methods provide accurate results, they are often resource-intensive and time-consuming. Machine learning (ML) provides a data-driven approach to identifying patterns and making predictions, enabling researchers to address complex problems that are challenging to solve through traditional methods [9]. For LiB cathode materials, ML has proven to be a promising tool in predicting critical material properties, including the crystal system, based on input features such as chemical composition and thermody-namic properties [10]. Researchers today have access to massive amounts of information about the predicted properties of materials. For example, the Materials Project provides a free, web-based platform where anyone can explore the physical and chemical properties of both known and predicted materials. These properties are calculated using density functional theory (DFT), which helps estimate the physical properties such as structure and electronic band gap of materials [11]–[13]. Improvements in exchange-correlation potentials have made it possible to accurately calculate the physical properties of many different materials, including those used in lithium-ion batteries [14]– [17]. Phosphate-based cathode materials with Li–P–(V, Mn, Fe, Co, Ni)–O compositions are of great interest for research due to their high capacity (volumetric and gravimetric) and stability [18], [19]. This study utilizes various classification algorithms to predict the crystal systems (Monoclinic, Triclinic, and Orthorhombic) of cathode materials with Li–P–(V, Mn, Fe, Co, Ni)–O compositions based on data obtained from the Materials Project. Fig.1 explains schematically the machine learning framework for data processing, including model training and feature selection. Machine learning models, including Linear Discriminant Analysis (LDA), Support Vector Machines (SVM), K-Nearest Neighbours (KNN), and ensemble methods such as Random Forest (RF), Extremely Randomized Trees (ERT), and Gradient Boosting Machines (GBM), were employed for this task. Monte Carlo Cross-Validation (MCCV) was conducted to ensure reliable performance evaluation. A sequential-forward-selection (SFS) approach was employed to enhance model interpretability and performance and identify each machine learning model's top three most influential features. These key features were used to construct feature subsets, and models were retrained and evaluated with different feature combinations. This approach identified the optimal (minimal) feature sets for achieving high prediction accuracy and provided valuable insights into the role of specific cathode properties in influencing the crystal system.

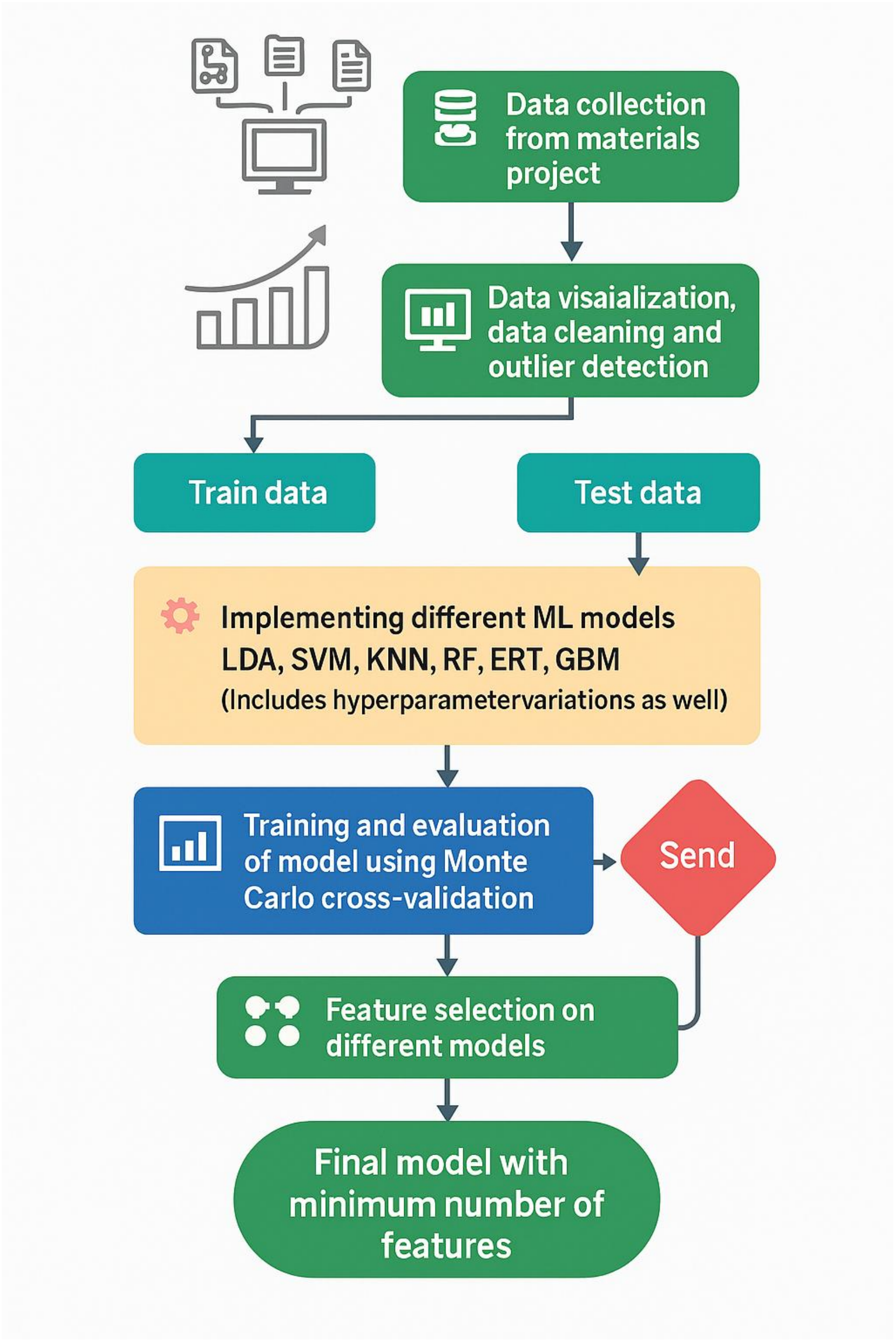

Figure 1. The schematic view of the Machine learning Framework for Data Processing, Model Training, and Feature Selection

## II. METHODOLOGY

### A. The dataset

The dataset consists of results from DFT calculations for 1819 cathode materials with Li–P–(V, Mn, Fe, Co, Ni)–O compositions derived from the Materials Project. The Materials Project generates data using advanced computational tools and methods, leveraging density functional theory (DFT) framework using a generalized gradient approximation (GGA) functional parametrized by Perdew, Burke, and Ernzerhof (PBE) [20], [21]. The transition metals, Fe, Co, Mn, Ni, and V, have been assigned a U parameter to correct for the self-interaction error present in GGA [22], [23]. The DFT calculations and optimizations in the Materials Project are carried out using VASP software [24]. The selected materials are symbolically presented in Table I together with the respective material parameters used in the present work. The complete data set is provided in the supplementary information. The dataset includes attributes such as chemical formula, space group symbol, space group number, formation energy ($E_f$), energy above hull ($E_H$), bandgap ($E_g$), number of atomic sites ($N_s$), density ($\rho$), unit cell volume (V), and crystal system. The parameters $N_s$ and $\rho$ represent the number of atoms in a unit cell and the density of the bulk crystalline material, respectively. Machine learning models in this context often use $V=M/\rho$, (where M is formula unit mass) as a primary variable. $E_H$ describes the energy required for decomposition into the most stable phases [13].

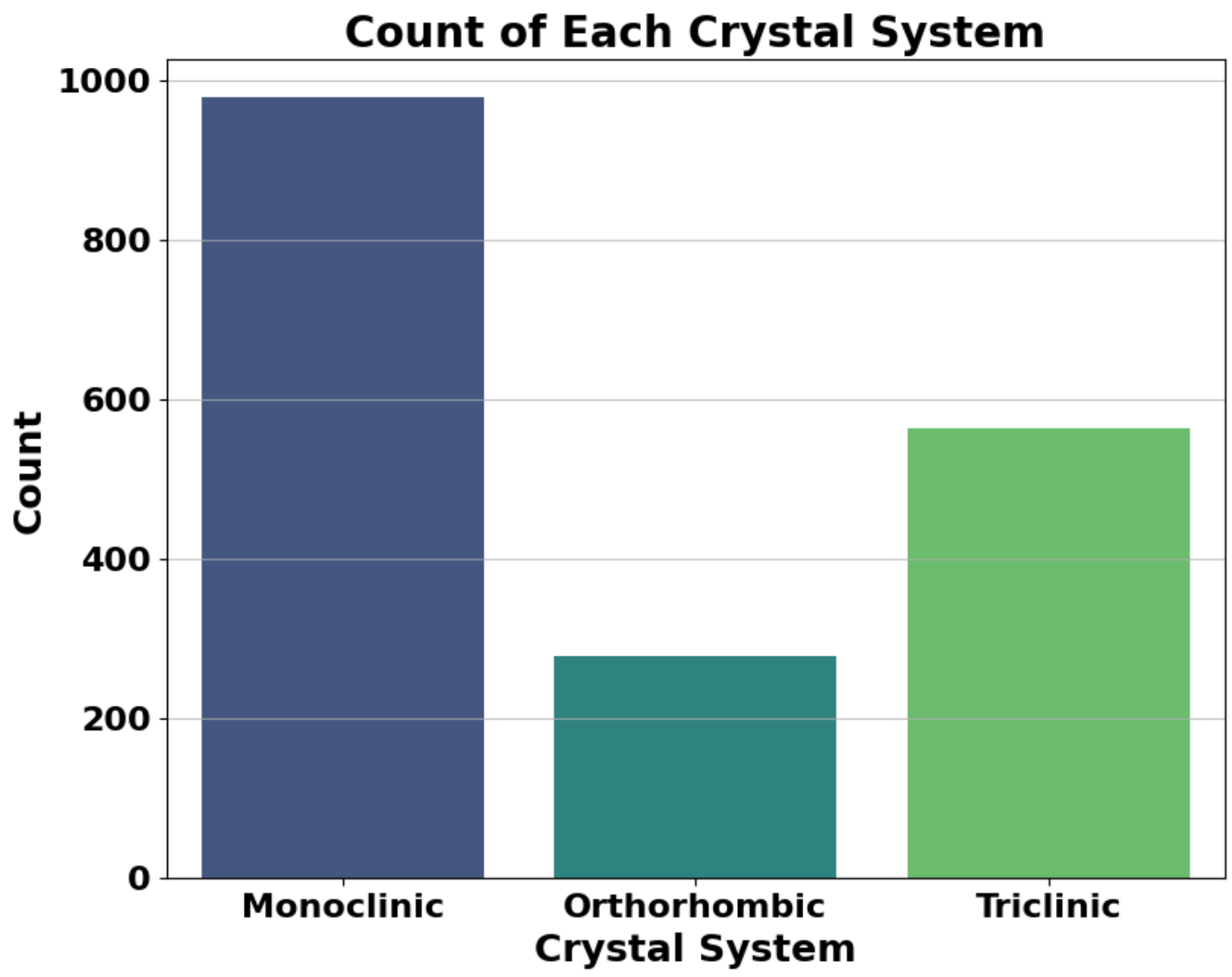

Figure 2. Bar plot showing the distribution of three different crystal systems

All properties in the dataset are calculated at 0 K and ambient pressure, with $E_g$ and V potentially influenced by temperature and pressure. However, these factors are held constant for this analysis. Fig.2 shows that among the 1,819 data points, Monoclinic, Triclinic, and Orthorhombic crystal systems contain 978, 564, and 277 data points, respectively. Formation energy, energy above the hull, bandgap, sites, and volume were used as input features in the machine learning model, with the crystal system serving as the output feature.

**B. Dataset Statistical Analysis**

A pair plot, Fig. 3, is a valuable tool for visualizing the relationships between multiple features in a dataset. It creates a matrix of scatter plots, each representing the relationship between two features. The diagonal of the matrix typically shows the distribution (density plots) of individual features. In a classification problem, a pair plot shows how well the different classes are separated based on the features. The box plots, Fig. 4(a-e), illustrate the distribution of key input features (Volume, Sites, Energy Above Hull, Formation Energy, and Band Gap) across three crystal systems: Monoclinic, Orthorhombic, and Triclinic. These plots provide valuable insights into the distribution of features within the different crystal systems, enabling the identification of trends, variability, and outliers, which are crucial for feature analysis and model refinement. Monoclinic systems generally show the widest ranges and more outliers in properties like Band Gap, Volume, and Formation Energy. Triclinic systems tend to have narrower ranges but more extreme outliers in Energy Above Hull. Orthorhombic systems exhibit intermediate distributions with fewer outliers.

**TABLE I:** Dataset of selected phosphate materials with properties used in this study

| Formula | Space Group Symbol | Space Group Number | Sites (Ns) | EH (eV) | EF (eV) | Volume (Å$^3$) | Density (g/cm$^3$) | Band Gap (eV) | Crystal System |
|---|---|---|---|---|---|---|---|---|---|
| $Li_2FeP_2O_7$ | P-1 | 2 | 48 | 0.033236 | -2.539132 | 587.586162 | 2.754 | 4.3546 | Triclinic |
| $Li_2VP_4O_{13}$ | P-1 | 2 | 40 | 0.045837 | -2.616142 | 537.320824 | 2.452 | 1.8286 | Triclinic |
| $Li_6Ni_5(P_2O_7)_4$ | P-1 | 2 | 94 | 0.038332 | -2.407053 | 1049.598321 | 3.261 | 3.4812 | Triclinic |
| $Li_2V_3(PO_4)_3$ | C2 | 5 | 40 | 0.059565 | -2.599565 | 474.363850 | 3.162 | 0.4490 | Monoclinic |
| $Li_3Mn_2(PO_3)_7$ | P-1 | 2 | 66 | 0.083130 | -2.558361 | 846.570725 | 2.681 | 4.0116 | Triclinic |
| . | . | . | . | . | . | . | . | . | . |
| . | . | . | . | . | . | . | . | . | . |
| . | . | . | . | . | . | . | . | . | . |
| $LiCo(PO_3)_3$ | P212121 | 19 | 56 | 0.000464 | -2.498897 | 640.941708 | 3.138 | 2.7884 | Orthorhombic |

Overall, monoclinic systems display more significant variability than the other two. The Heatmap, Fig 4(f), showcases the linear relationships between the numerical features in the dataset. A strong negative correlation exists between Formation Energy and the band gap ( -0.15). This suggests that materials with low formation energy have a higher band gap. Positive correlations, such as between Sites and Volume (0.98), indicate that larger unit cells (more sites) tend to have higher volumes, and Weak correlations (-0.0059) imply negligible relationships between specific features, like energy above hull and volume.

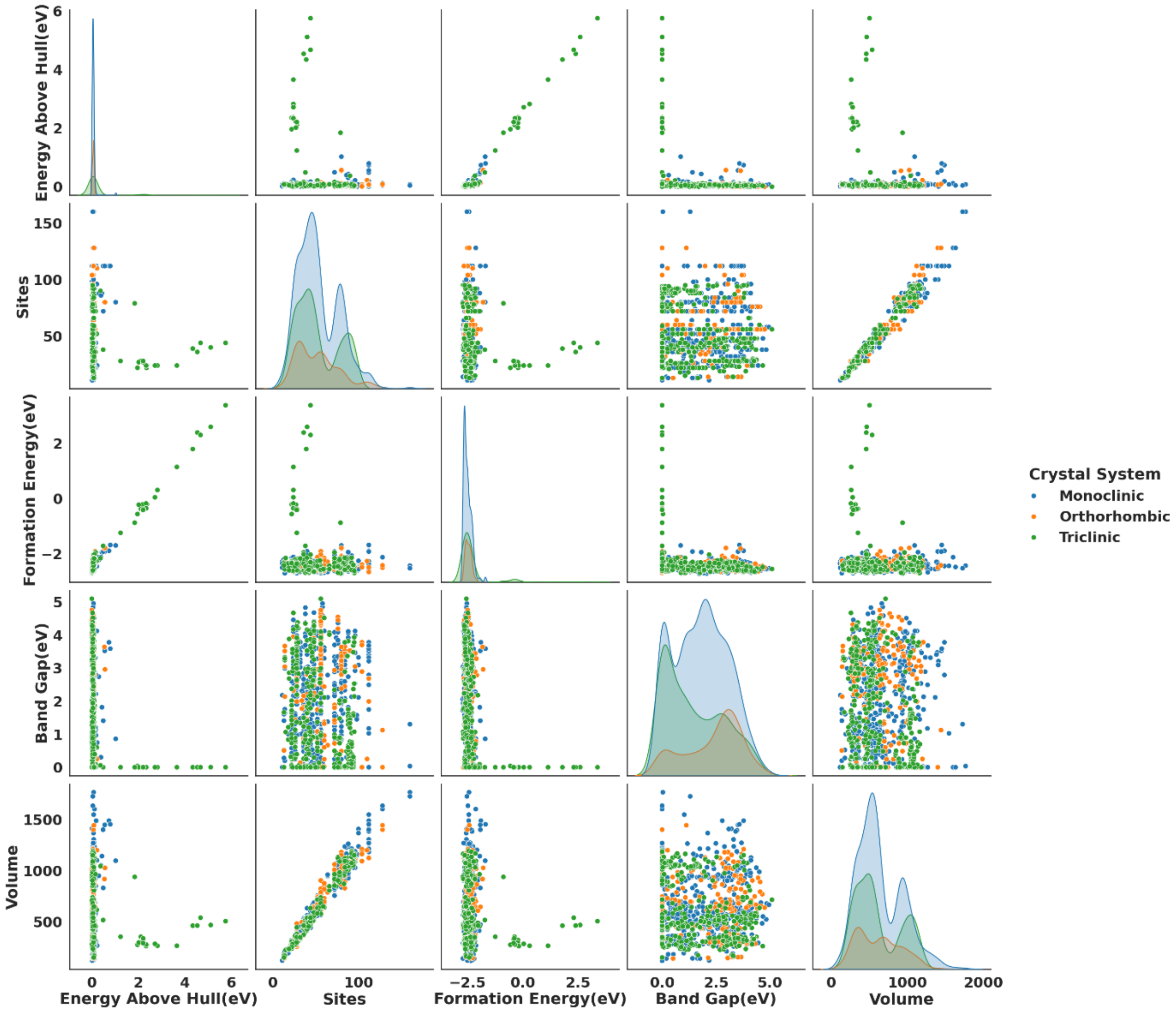

Figure 3. The pairs plot for different properties of Li– (Mn, Fe, Co, Ni, V)–P–O cathodes based on the extracted data from the Materials Project.

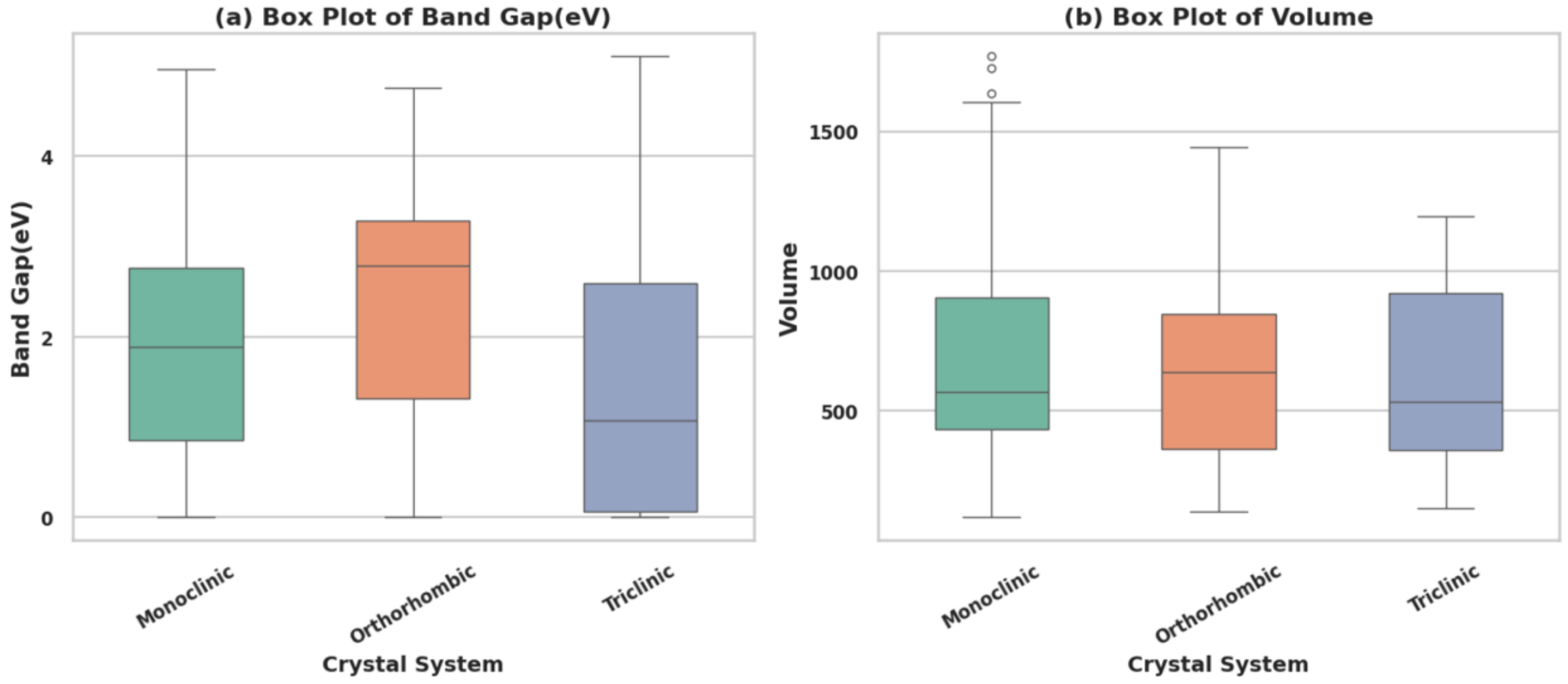

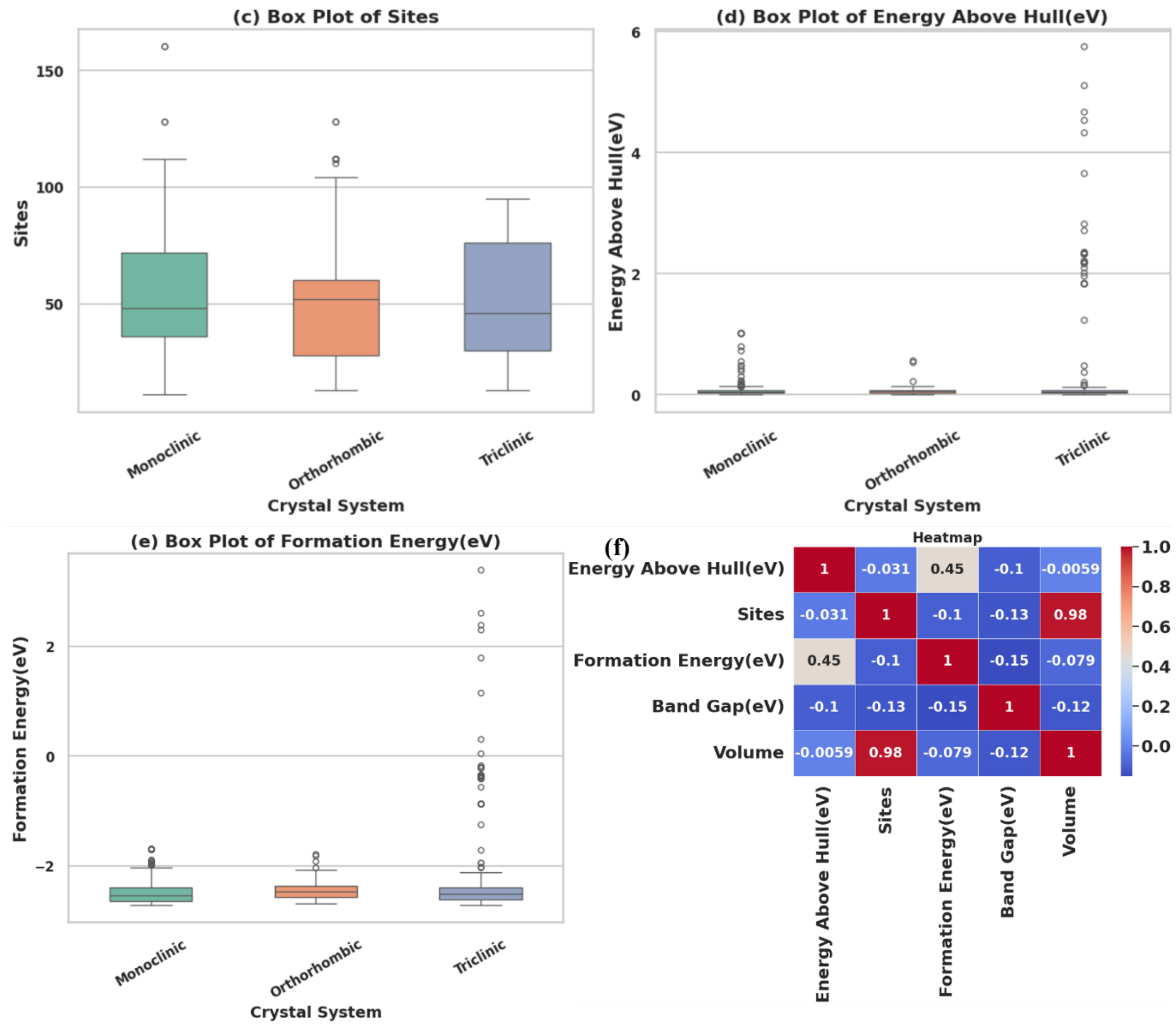

Figure 4. (a-e) Box plot and (f) heatmap for correlation analysis among different features.

## C. Outlier Detection and Removal

Outlier detection was performed to identify extreme values within the features of the dataset, and the results are summarized in Fig.5. The Interquartile Range (IQR) method, a robust statistical approach, effectively identified and removed data points falling outside a reasonable range. The IQR was calculated as the difference between each feature's 75th percentile (Q3) and 25th percentile (Q1). Outliers were defined as values below (Q1 - 1.5 × IQR) or above (Q3 + 1.5 × IQR).

The dataset contained 1,819 data points. After applying the IQR-based removal process, approximately 4.84% of the total data points were identified as outliers, reducing the dataset to 1,731 data points.

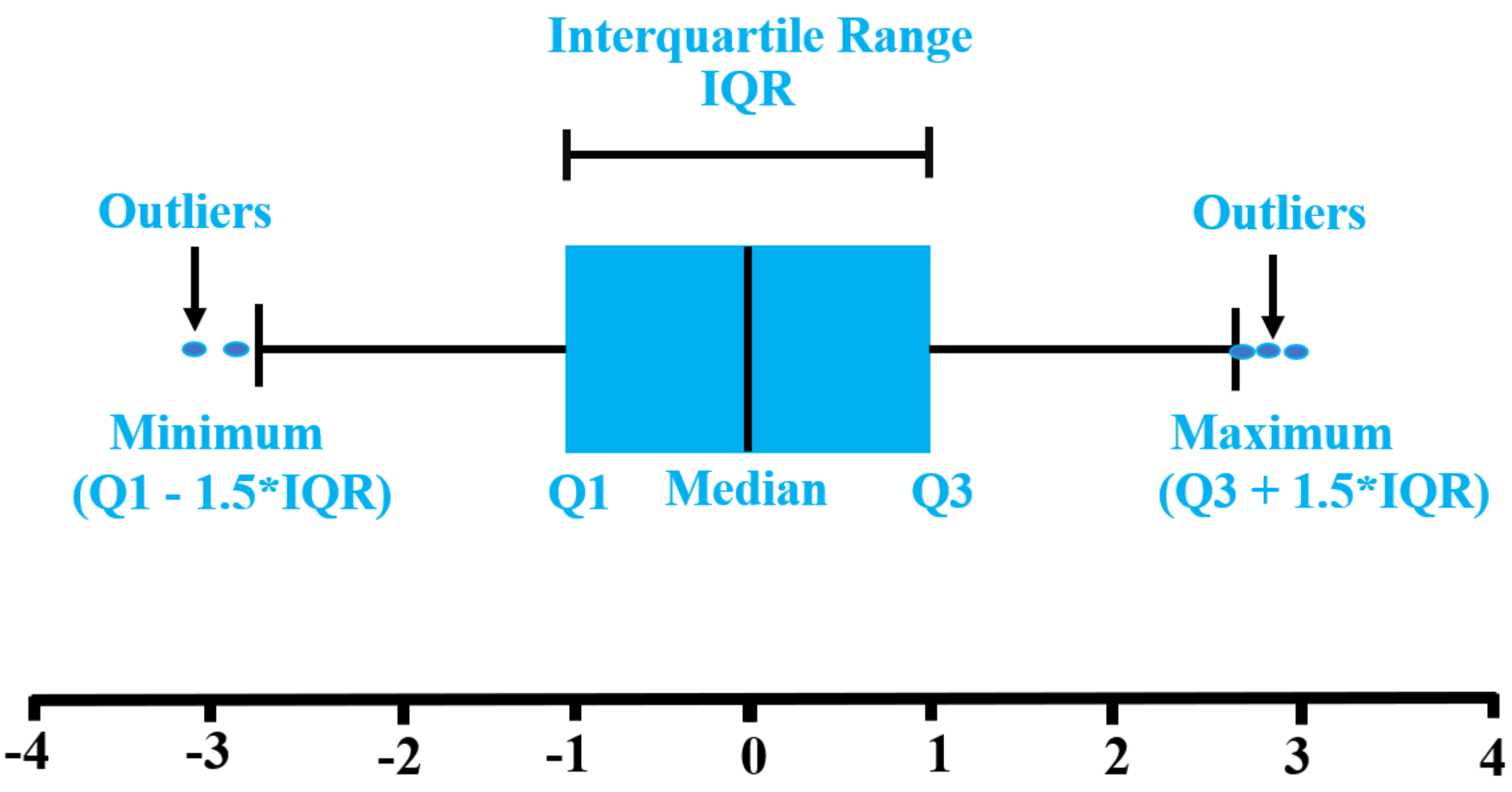

Figure 5. Schematics of Outlier Detection and Elimination via Interquartile Range

**D. CLASSIFICATION TECHNIQUES**

**1) Linear Discriminant Analysis:** Linear Discriminant Analysis (LDA), Fig. 6(a), is a supervised dimensionality reduction technique used for classification. It projects high-dimensional data onto a lower-dimensional space while maximizing class separability [25]. LDA assumes that class distributions are Gaussian with a standard covariance matrix [26].

The method involves computing class means, within-class ($S_W$) and between-class ($S_B$) scatter matrices, and solving the eigenvalue problem, i.e., $S_B\, w = \lambda\, S_W\, w$; to obtain optimal projection directions [27]. For a 3(or k)-class problem, data is reduced to 2(or k-1) dimensions. The transformed data is classified using class centroids and distance metrics.

**2) Support Vector Machine (SVM):** Support Vector Machine (SVM), Fig. 6(b), is a supervised learning algorithm widely used for classification tasks due to its ability to find an optimal hyperplane that maximizes the margin between different classes [28]. It is particularly effective in handling high-dimensional data and can be extended to non-linearly separable cases using kernel functions [29]. In this study, the Radial Basis Function (RBF) kernel was used to capture complex decision boundaries. The regularization parameter $C = 10$ was chosen to control the trade-off between maximizing the margin and minimizing classification errors.

**3) K-Nearest Neighbours (KNN):** K-Nearest Neighbours (KNN), Fig.6(c), is a simple, non-parametric classification algorithm that assigns a class label based on the majority class among its K nearest neighbours [30]. The model's performance depends on the choice of K and the distance metric [31].

For this study, we set K =3, meaning classification was based on the three closest neighbours. The Euclidean distance metric was used to measure proximity, and uniform weighting was applied to all neighbours. These hyperparameter choices ensured a balance between model complexity and classification accuracy.

**4) Random Forest:** Random Forest (RF), Fig. 6(d), is a popular and powerful ensemble-supervised classification method [32]. Random Forest (RF) is an ensemble classification algorithm that constructs decision trees using random subsets of data and features (bagging and random subspaces). Random selection of features at each split reduces tree correlation, improving prediction power and efficiency. RF overcomes overfitting, is less sensitive to outliers, and does not require pruning. It handles continuous, categorical data, manages missing values, and provides automatic measures of variable importance and accuracy. Combining bootstrapping, ensemble strategies, and random sampling ensures high prediction accuracy, generalizability, and interpretability, making RF suitable for high-dimensional datasets [33]. In this study, we used 100 estimators, a maximum tree depth of 10 to prevent overfitting, and a random state of 40 to ensure reproducibility.

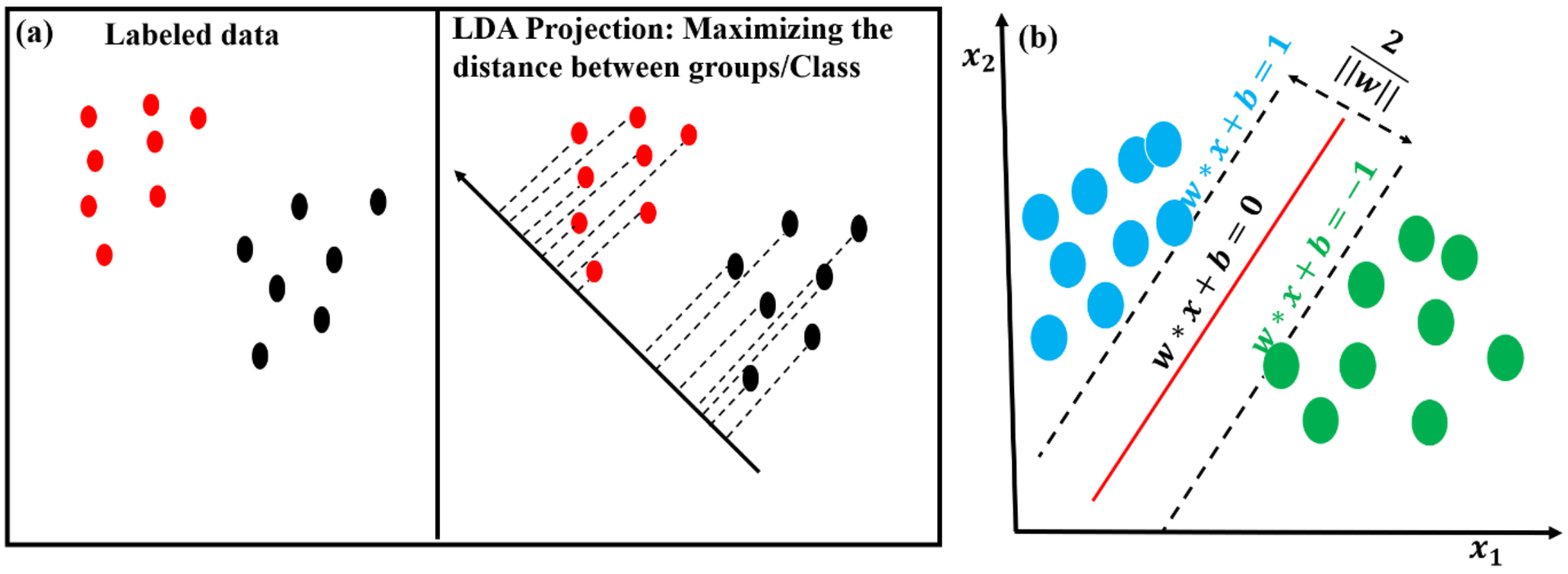

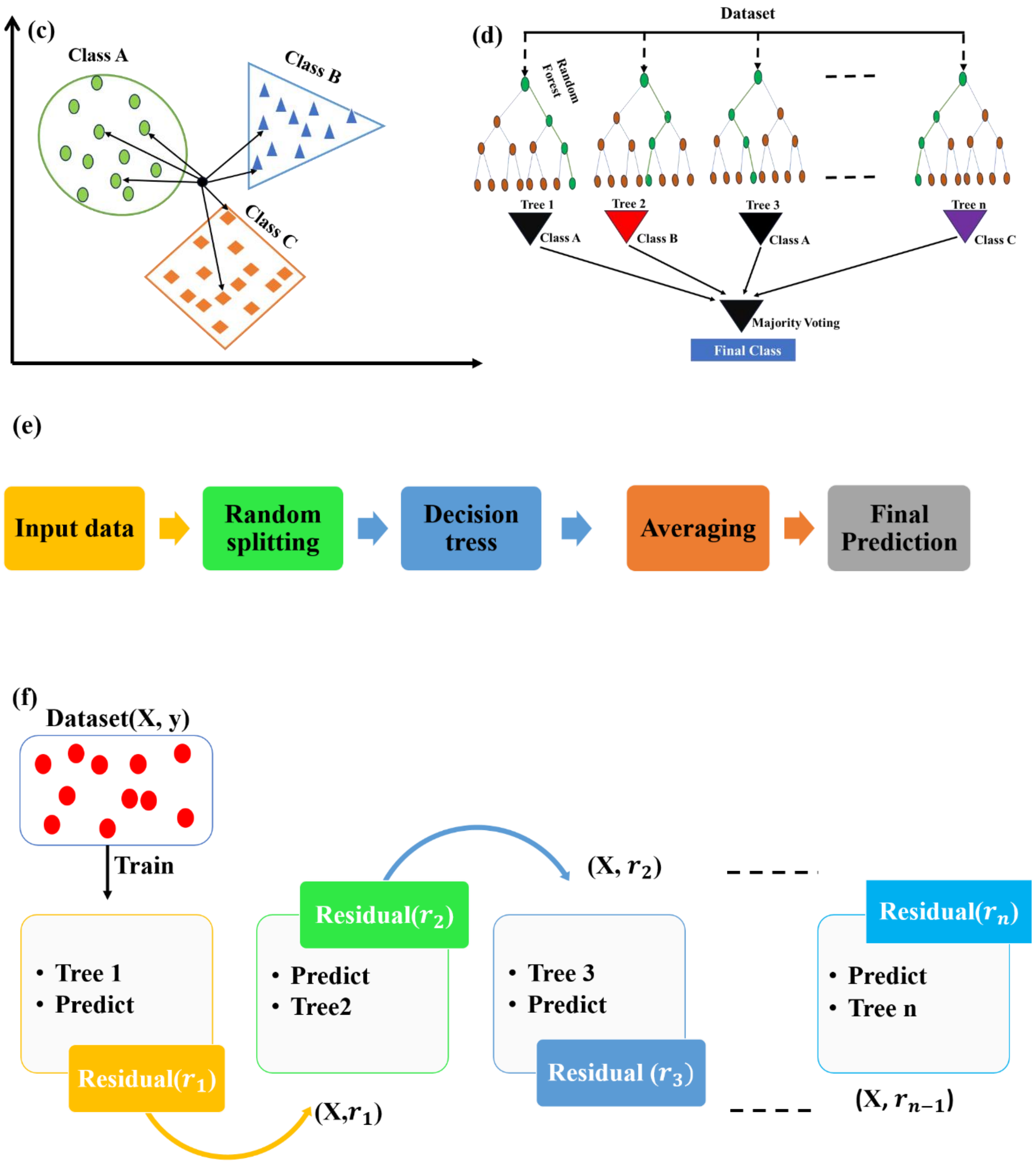

Figure 6. Schematic Representation of Machine Learning Models including (a) Linear Discriminant Analysis (LDA) for classification and dimensionality reduction, (b) Support Vector Machine (SVM) for optimal hyperplane-based separation, (c) k-Nearest Neighbors (K-NN) for instance-based classification, (d) Random Forest (RF) for ensemble-based decision-making, (e) Extremely Randomized Trees (Extra Trees) for enhanced randomness in tree con-struction, and (f) Gradient Boosting Machine (GBM) for iterative error minimization and improved predictive performance.

**5) Extremely Randomized Trees:** Extremely Randomized Trees (ERT), Fig. 6(e), is an ensemble learning method designed for classification and regression tasks. It enhances the Random Forest approach by introducing additional randomness during the construction of decision trees. Unlike traditional methods, ERT selects both the attributes and the splitting thresholds randomly rather than optimizing them. This approach cre-ates greater diversity among the trees in the ensemble, im-proving generalization and reducing the risk of overfitting.

The algorithm builds multiple unpruned decision trees, and predictions are made by averaging in regression tasks or through majority voting in classification problems. Due to its randomized splitting criterion, ERT is computationally efficient and achieves competitive accuracy, making it well-suited for high-dimensional datasets [34]. In this study, the model was configured with 150 estimators, a maximum tree depth of 20, and a random state of 40.

**6) Gradient Boosting Machine (GBM):** Gradient Boosting, Fig 6(f), is an ensemble learning technique that builds a strong predictive model by sequentially combining multiple weak models, typically decision trees [35]. It works by iteratively minimizing residual errors, where each new tree corrects the mistakes of the previous ones [36]. The algorithm follows steps as (i) a base model (typically a decision tree) is initialized to make initial predictions; (ii) the residuals (difference between actual and predicted values) are computed; (iii) a new decision tree is trained to predict these residuals; (iv) the model is updated by adjusting predictions with a scaled contribution from the new tree, controlled by the learning rate; and (v) this process is repeated for a predefined number of iterations to refine the predictions. For this study, the best hyperparameters considered are (i) Number of estimators = 200, (ii) Learning rate = 0.1, and (iii) Maximum depth = 5, to ensure optimal model performance.

## III. RESULT & DISCUSSION

This study used Python and the required libraries to build machine-learning models and calculate their performance. The dataset was divided into 80% for training and 20% for testing. The 80% training data was utilized to train multiple machine learning models, including LDA, SVM, KNN, RF, ERT, and GBM, followed by hyperparameter tuning. Among these models, the GBM, ERT, and RF models achieved the highest test accuracy: 76.95%,74.93%, and 76.37%, respectively, on the 20% test data because GBM, ERT, and RF are ensemble models; they outperform simpler models like LDA by captur-ing nonlinear relationships and feature interactions in the data. Unlike LDA, which assumes linear boundaries and Gaussian distributions, these models make no such assumptions and are robust to noise and outliers. Their ability to aggregate predictions from multiple decision trees enhances general-ization and accuracy, making them particularly effective for complex and high-dimensional datasets. Fig.7 summarizes the performance of the machine learning models by classifying data into three classes (0, 1, and 2) using confusion matrices. These matrices illustrate the alignment between predicted and actual class labels, providing a detailed classification accuracy assessment for each class. The labels 0, 1, and 2 represent the monoclinic, triclinic, and orthorhombic crystal systems. Monte Carlo Cross-Validation (MCCV), also known as repeated random sub-sampling, was utilized to assess model accuracy and detect overfitting. This method involves repeatedly splitting the dataset into training and randomly testing subsets. For each split, the model was trained on the training set and tested on the testing set, and the prediction accuracy was recorded. The process was repeated 100 times to ensure reliable and consistent results. The dataset was shuffled before each split using 100 predefined random seeds to standardize the evaluation, ensuring all classifiers were tested on identical splits for direct performance comparison.

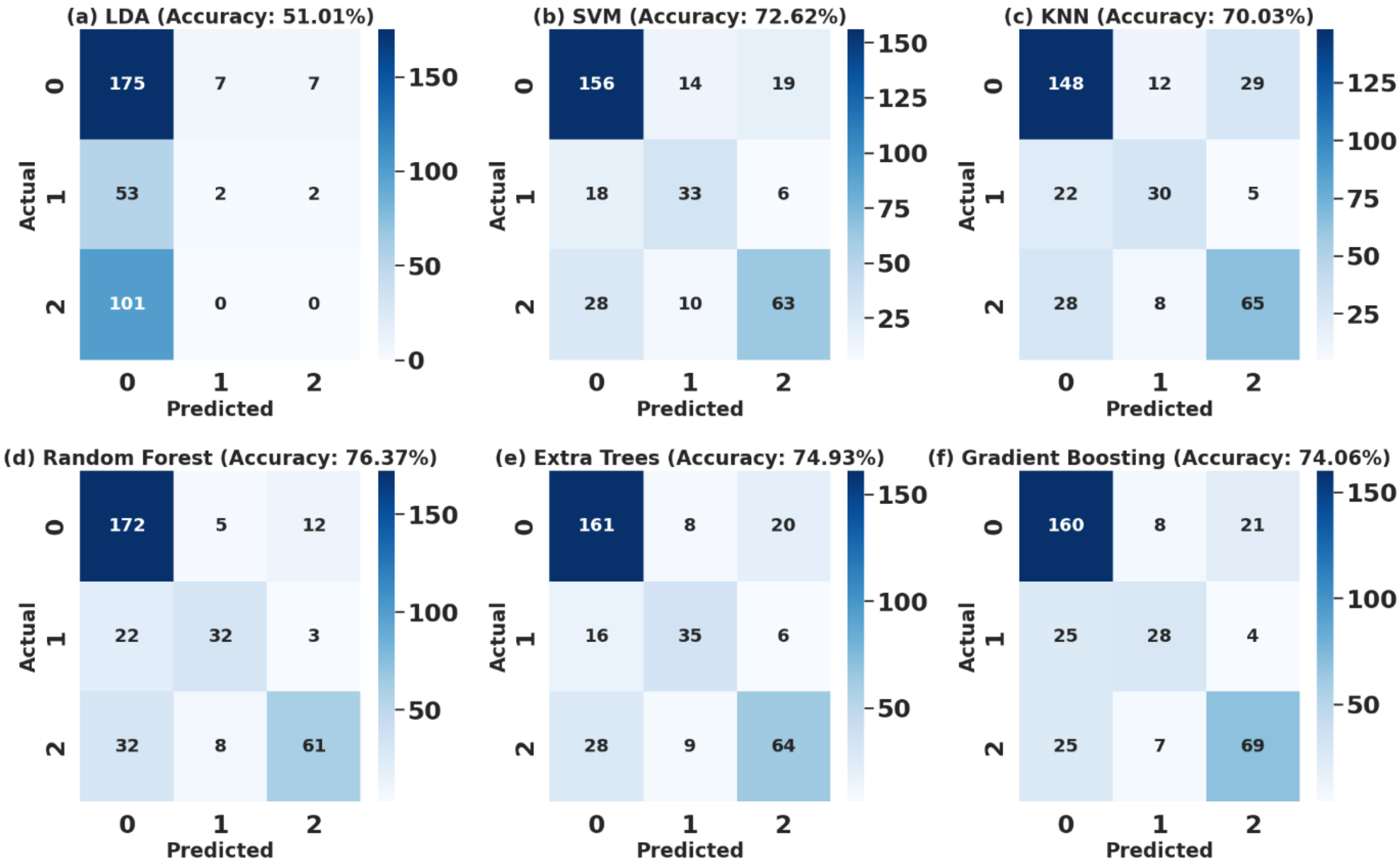

Figure 7. Confusion Matrix for the different models where 0, 1, and 2 correspond to monoclinic, triclinic, and Orthorhombic crystal systems, respectively.

The Fig.8 shows the relationship between training data percentage and mean accu-racy for six machine learning models using MCCV. GBM and ERT consistently achieve the highest accuracy, while Linear Discriminant Analysis (LDA) performs the worst. Increasing the training percentage improves accuracy for all models, but the gains diminish as it approaches 90%.

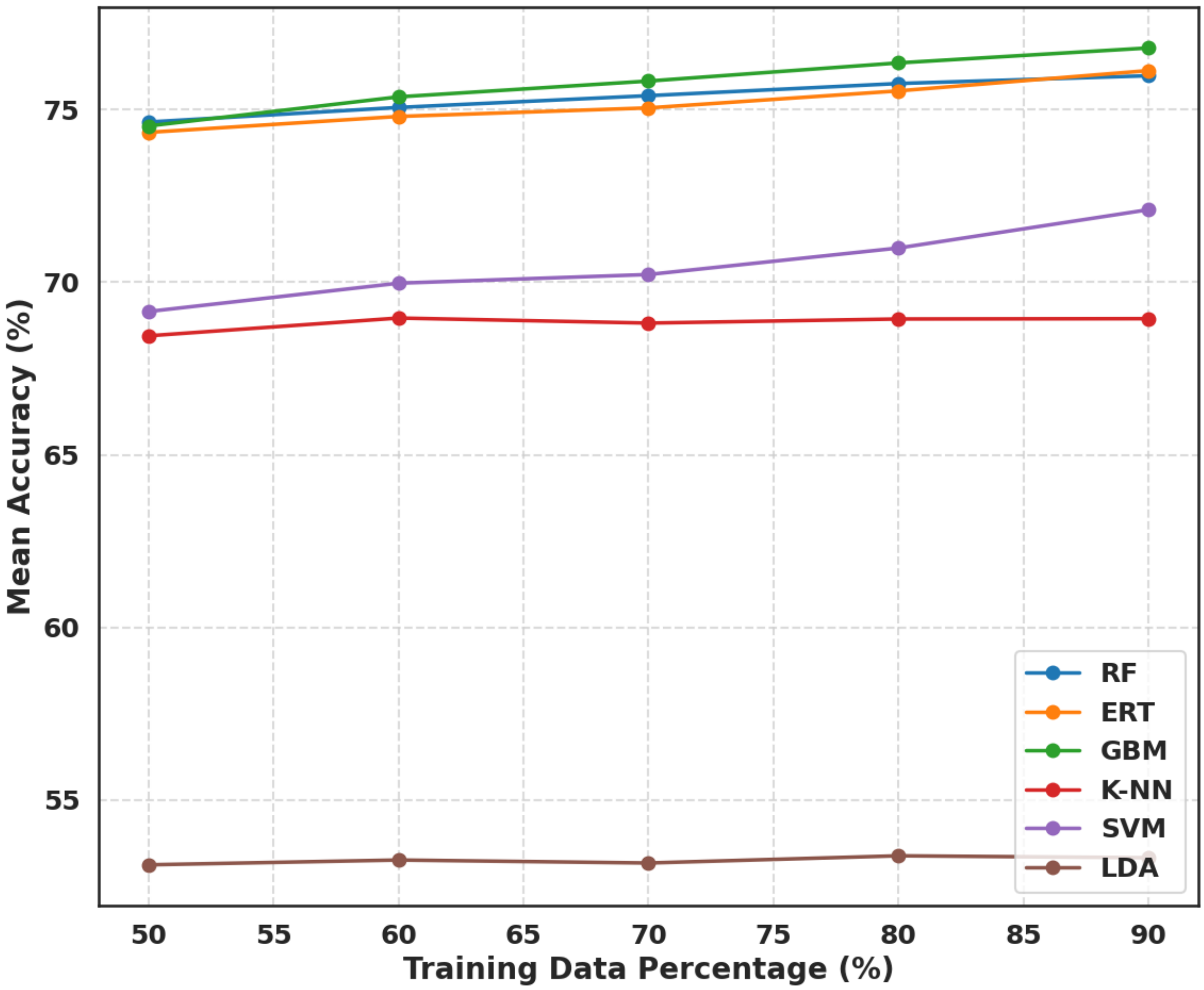

Figure 8. Effect of percentage of training data for building ML models on the mean accuracy.

Ensemble models like GBM and ERT exhibit robust performance and generalization compared to others, highlighting their effectiveness for the dataset, Fig 9. The plot shows that Extra Trees (ERT) and Gradient Boosting (GBM) exhibit the lowest standard deviation, indicating the most consistent and stable performance across different Monte Carlo Cross-Validation (MCCV) splits. This reflects their robustness to variations in train-test splits and their ability to generalize well to unseen data. Their ensemble-based mechanisms effectively minimize variability, making them reliable models for this task. Lower std for these mod-els implies higher dependability in predictive performance. Following Monte Carlo Cross-Validation, Fig.10 Sequential Forward Selection (SFS) enhanced model performance and interpretability by identifying the most significant input fea-tures. This iterative process starts with no features and adds the feature that improves model accuracy the most, continuing until a predefined number of features (three in this case) are selected. The models were retrained and evaluated using the selected features, reducing dimensionality and focusing on the most relevant properties. SFS highlighted the importance of features like volume, bandgap, and atomic sites in improving classification accuracy across all models. Once the top three features were identified, the classifier was trained using 80% of the data with only these selected features. The model's accuracy was then evaluated on the remaining 20% of the data. This approach reduces dimensionality, focuses on the most relevant features, and enhances the classifier's overall performance.

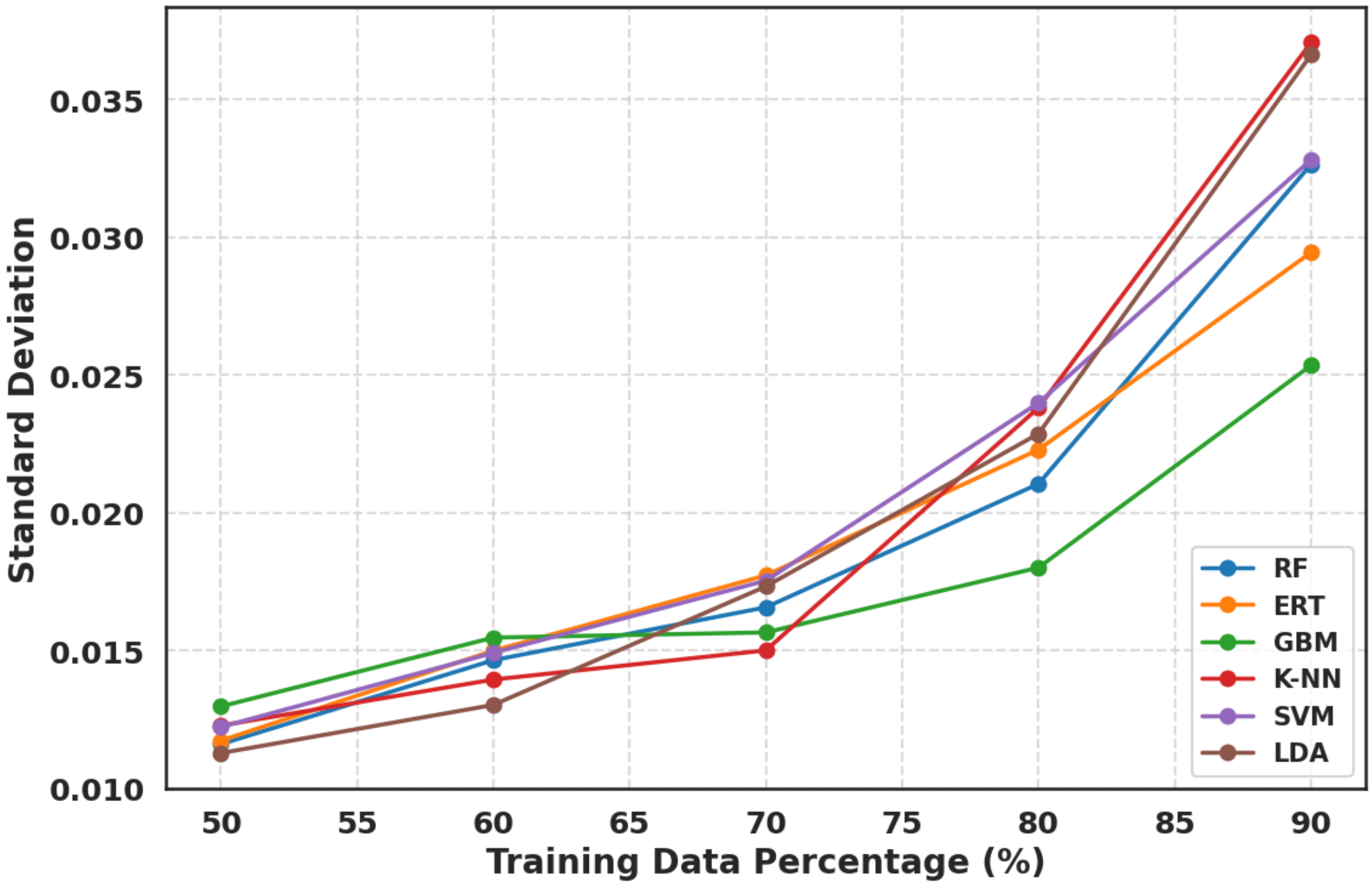

Figure 9. The curve between training data percentage Vs standard deviation in MCCV

Table II summarizes the results of Sequential Feature Selection applied to six machine learning models. The table identifies the top three features for each model, including Formation Energy $E_f$, Energy Above Hull $E_H$, Volume, Band Gap, and Sites, along with their corresponding accuracies. GBM achieved the highest accuracy (80.40%) using Volume, Band Gap, and Sites, while Sites were consistently selected across all models. These results demonstrate the effectiveness of SFS in reducing dimensionality and improving model performance by focusing on the most relevant features.

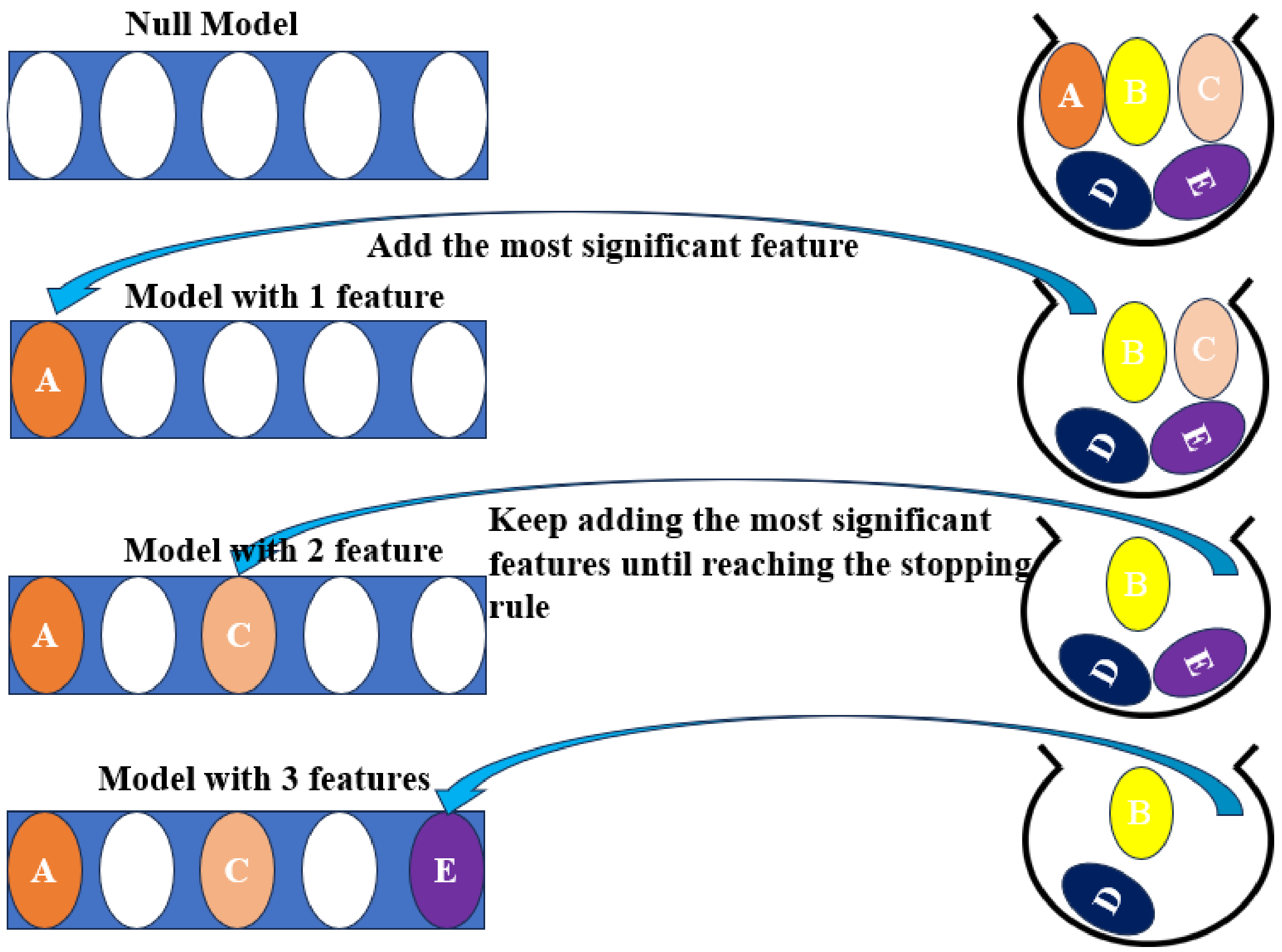

Figure 10. Schemetic representation of forward sequential feature selection

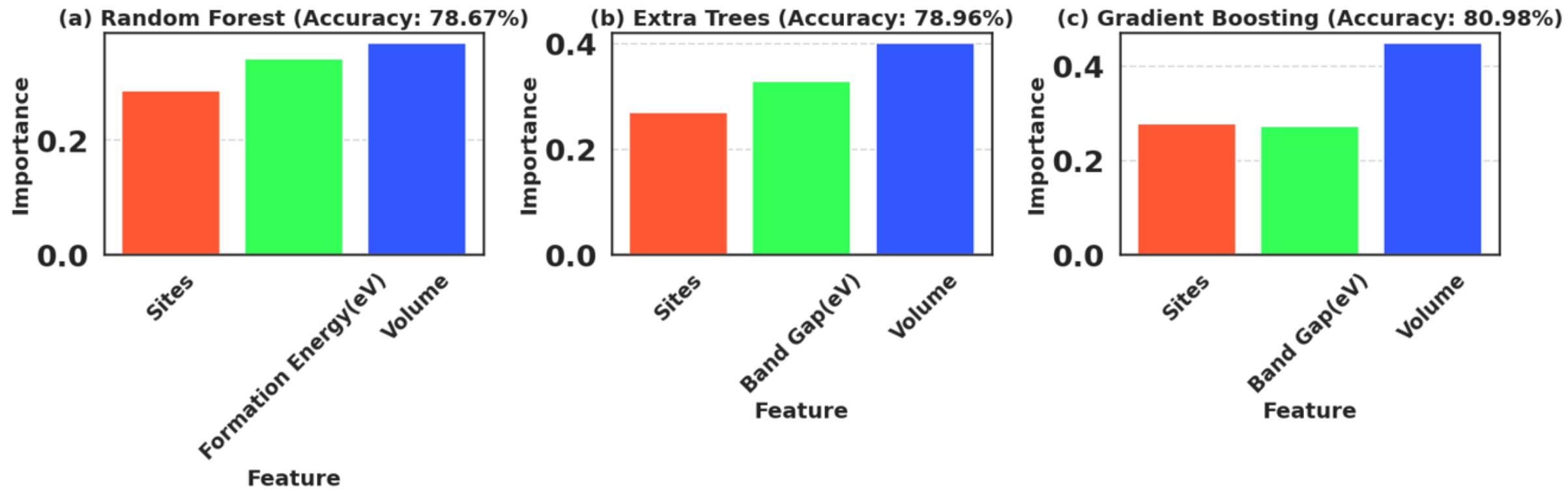

Figure 11. Top Features and Their Importance for Best-Performing Models

Fig.11 illustrates the performance of the best ensemble machine learning models, highlighting their top three input features based on importance, along with the corresponding accuracy achieved using these features.

TABLE II: Machine learning models with their selected features and corresponding accuracy

| **Model** | **$E_f$** | **$E_H$** | **Volume** | **Band Gap** | **Site** | **Accuracy (%)** |
|---|---|---|---|---|---|---|
| LDA | | √ | √ | | √ | 56.48 |
| SVM | | √ | | √ | √ | 66.28 |
| KNN | √ | | √ | | √ | 70.32 |
| RF | √ | | √ | | √ | 78.67 |
| ERT | | | √ | √ | √ | 78.96 |
| GBM | | | √ | √ | √ | 80.40 |

Table III represents the results of training six machine learning models (LDA, SVM, KNN, RF, ERT, GBM) using different input feature combinations identified through Sequential Feature Selection (SFS). Each row represents a unique combination of features, such as Energy Above Hull $E_H$, Formation Energy $E_f$, Volume, Band Gap, and Sites, and their corresponding accuracies. Models like GBM and RF con-sistently achieved high accuracy across combinations, show-casing their robustness, while LDA performed poorly, Fig. 12.

TABLE III: Accuracy (%) of a machine learning model with different input features.

| **Features** | **LDA** | **SVM** | **KNN** | **RF** | **ERT** | **GBM** |
|---|---|---|---|---|---|---|
| $E_H$, Band Gap, Sites | 57.06 | 66.28 | 68.88 | 74.93 | 74.06 | 72.91 |
| $E_f$, Sites, Volume | 56.48 | 68.30 | 73.49 | 78.67 | 77.52 | 79.54 |
| $E_H$, Sites, Volume | 56.48 | 71.47 | 71.47 | 76.95 | 77.81 | 80.40 |
| Band Gap, Sites, Volume | 56.48 | 64.84 | 75.50 | 80.69 | 78.96 | 80.40 |

The combination of Band Gap, Sites, and Volume as input features yielded the best overall performance, with RF achieving 80.69% and GBM 80.40% test accuracy. This highlights the importance of feature selection in enhancing model perfor-mance. In this study, we assessed the thermodynamic stability of cathode materials using a filtering process, focusing on energy above hull and formation energy as stability indicators.

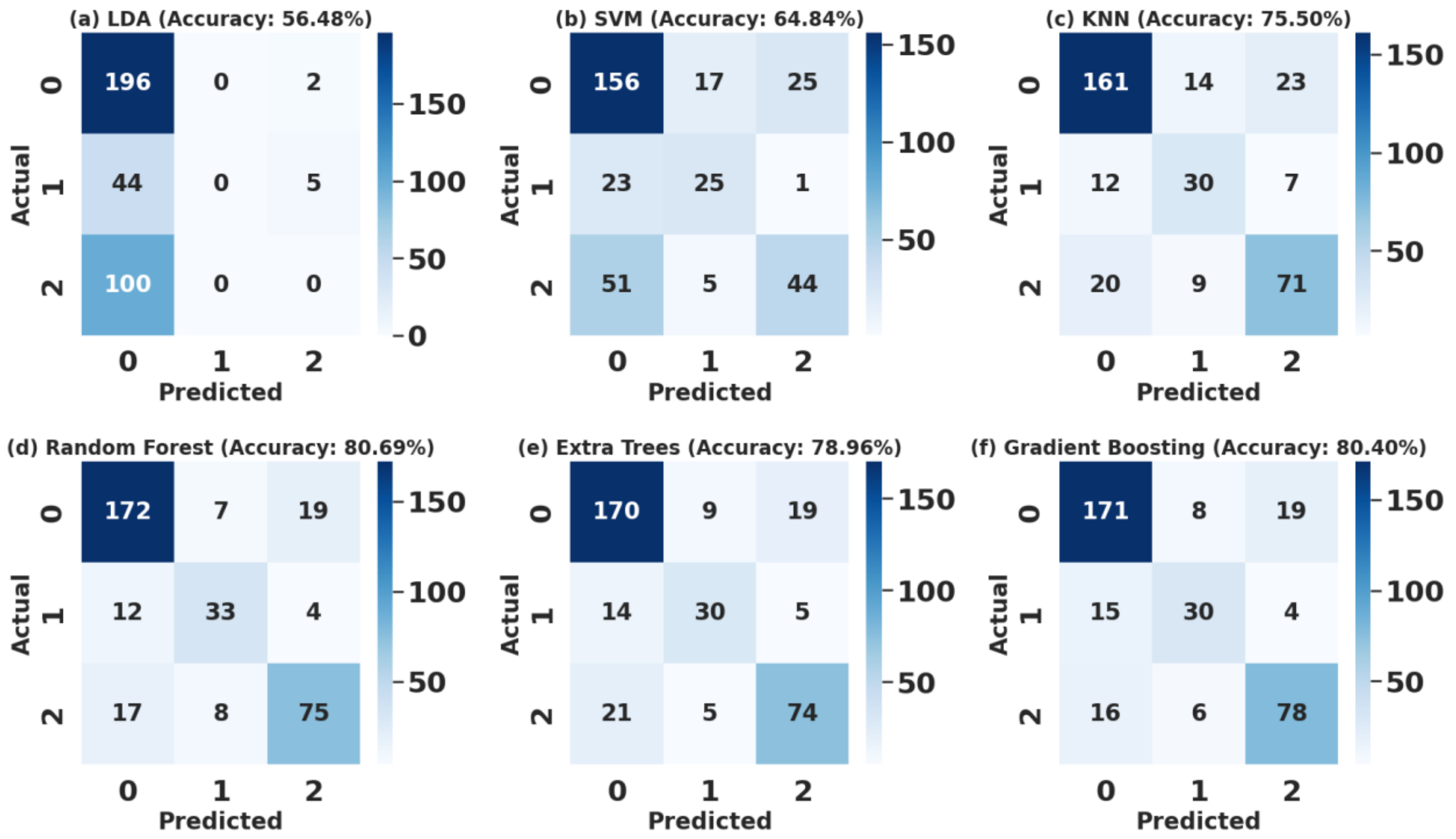

Figure 12. The performance matrix of different models with Band Gap, Sites, and Volume as input features.

Materials were filtered based on two criteria: energy above hull = 0.014 eV and formation energy < -2.672 eV. The analysis revealed that vanadium-based materials, particularly $Li_3V_2(PO_4)_3$ in the monoclinic phase, exhibited the highest stability, with a formation energy of -2.725 eV and energy above hull of 0 eV. These results suggest $Li_3V_2(PO_4)_3$ as a promising candidate for lithium-ion battery cathodes.

## IV. CONCLUSION

This study demonstrated the effectiveness of machine learning models in classifying Li–P–(V, Mn, Fe, Co, Ni)–O based cathode materials into three crystal systems (i.e., monoclinic, triclinic, and orthorhombic crystal structures). Sequential Feature Selection (SFS) and Monte Carlo Cross-Validation (MCCV) are used as key methods in optimizing model performance and ensuring robust evaluation. The ensemble models like GBM, ERT, and RF achieved the highest accuracy: 80.40%, 78.96%, and 80.69%, respectively, using the top three selected features: Volume, Band Gap, and Sites. This study highlights the power of ensemble methods and optimized feature selection for achieving reliable and robust predictions in crystal system classification tasks. Apart from this, the V-based phosphate-based materials are more stable than others.

## ACKNOWLEDGMENT:

Ambesh Dixit acknowledges CRG, DST-SERB, Government of India, through project number SERB/F/10090/2021-2022, for financial assistance to carry out this work. Authors also acknowledge the Indian Institute of Technology Jodhpur, India, for providing high-performance computing clusters for the present work.